\documentstyle[prd,tighten,aps]{revtex}
\begin{document}
\draft
\twocolumn[\hsize\textwidth\columnwidth\hsize\csname
@twocolumnfalse\endcsname

\title{Quantization of pure gravitational plane waves}
\author{Guillermo A. Mena Marug\'an} 
\address{I.M.A.F.F., C.S.I.C., Serrano 121, 28006 Madrid, Spain} 
\author{Manuel Montejo}
\address{Universitat de Barcelona, Diagonal 647, 08028
Barcelona, Spain}
\date{25 June 1998}
\maketitle

\vspace*{-4.5cm}\begin{flushright}
gr-qc/9806105
\end{flushright}\vspace*{3.3cm}

\begin{abstract}
Pure gravitational plane waves are considered as a special case of
spacetimes with two commuting spacelike Killing vector fields. 
Starting with a midisuperspace that describes this kind of 
spacetimes, we introduce gauge-fixing and symmetry conditions
that remove all non-physical degrees of freedom and ensure that 
the classical solutions are plane waves. In this 
way, we arrive at a reduced model with no constraints and whose only
degrees of freedom are given by two fields. In a suitable coordinate
system, the reduced Hamiltonian that generates the time evolution 
of this model turns out to vanish, so that all relevant information 
is contained in the symplectic structure. We calculate this 
symplectic structure and particularize our discussion to the case
of linearly polarized plane waves. The reduced phase space can then 
be described by an infinite set of annihilation and creation like 
variables. We finally quantize the linearly polarized model by 
introducing a Fock representation for these variables.
\end{abstract}

\vskip2pc]

\renewcommand{\thesection}{\Roman{section}}
\renewcommand{\theequation}{\arabic{section}.\arabic{equation}}

\section{Introduction}

The analysis of gravitational waves has received lately a great deal 
of attention, both from the theoretical and the experimental points 
of view\cite{EXP}. In particular, there is an increasing 
interest in the study and quantization of spacetimes with two 
commuting Killing vectors that describe wave solutions in 
source-free general relativity [2-5]. One of the reasons for this 
interest, apart from possible applications in cosmology and 
astrophysics, is that this kind of spacetimes provides a good arena 
to test quantization techniques and discuss conceptual issues in 
gravity. The symmetry reduction of Einstein gravity in the presence
of two commuting Killing vectors leads in general to models with 
an infinite number of degrees of freedom, i.e., to midisuperspace 
models. It is commonly accepted that the quantization of these 
models, which is given by a true quantum field theory, may be 
significant for discussing basic features of the outstanding
theory of quantum gravity. In contrast, minisuperspace truncations 
of general relativity, like, e.g., most of the gravitational systems
quantized in the literature\cite{MINI}, are too simple to capture 
the field complexity that should be present in full quantum gravity.
Another motivation for studying the reduction of general relativity
by two commuting spacelike Killing vectors is that it 
admits a similar formulation to that for coset space $\sigma$-models 
coupled to two-dimensional gravity and a dilaton\cite{KOR}. The 
quantization of any of these two types of systems should hence allow
a better understanding of the quantum physics of the other.

The spacetimes with two commuting Killing vectors that have been 
quantized so far are the family of cylindrically symmetric 
gravitational waves and the Gowdy model with the spatial topology of
a three-torus. The Gowdy universes are vacuum spacetimes with compact
sections of constant time\cite{GO}. They can be thought of as 
inhomogeneous spacetimes filled with gravitational waves.
The quantization of the Gowdy model when the topology of the 
sections of constant time is that of a three-torus was analyzed in 
Ref.\cite{ME}. Preliminary studies of this quantization had been 
carried out by Berger\cite{BE} (assuming that the Killing vectors of
the model were hypersurface orthogonal), and by Husain and 
Smolin\cite{VI}. On the other hand, the quantization of cylindrical 
gravitational waves with linear polarization was achieved by Ashtekar
and Pierri\cite{AP}, completing previous works on the subject by 
Kucha\v{r}\cite{K} and Allen\cite{AL}. The effects produced by 
quantum fluctuations in this cylindrically symmetric 
model were studied in Ref.\cite{ASH97}. Finally, the more general 
case of cylindrical waves with two polarizations was recently 
quantized by Korotkin and Samtleben\cite{KS}. 

In addition to possessing two commuting spacelike Killing vector 
fields, the above families of spacetimes satisfy another condition: 
the existence of two-surfaces that are everywhere orthogonal to the 
group orbits spanned by the Killing vectors. For spacetimes of this 
class, a specially useful metric function is the square root of 
the determinant of the two-metric that corresponds to the group 
orbits\cite{MAC}. We will denote this positive function by $W$. 
The spacelike, timelike, or null character of the gradient of $W$ is
invariant under coordinate transformations. In this paper, we will 
study the quantization of another example of gravitational waves
that belongs to this class of spacetimes, namely,
the case of pure gravitational plane waves\cite{MAC,VER}. Our 
analysis will exhaust the discussion of the different posibilities 
for the gradient of $W$, in the sense that this gradient is
timelike for the Gowdy model with the topology of a 
three-torus\cite{GO}, spacelike for cylindrical gravitational 
waves\cite{AP}, and null for plane waves. 

Gravitational plane waves are a special type of pp-waves, i.e.,
plane-fronted gravitational waves with parallel rays\cite{MAC,KUN}. 
They were first studied by Baldwin and Jeffrey\cite{BAL}, and admit 
a five-dimensional group of isometries with a three-dimensional 
Abelian subgroup that acts on null hypersurfaces\cite{BPR}. 
Gravitational plane waves can be interpreted as describing the 
gravitational field at great distances from finite radiating 
bodies\cite{MAC,BAL}. Some of these waves are related to certain 
limits of Bianchi cosmological models and inhomogeneous
generalizations of such models\cite{VER,BI}. On 
the other hand, the collision of two gravitational plane waves has 
received intensive study, and a lot of solutions that represent 
this collision are known\cite{VER,COL}. 

In this work, we will restrict our considerations to the case of 
pure gravitational plane waves, namely, plane waves that are 
solutions to Einstein equations in vacuum. Preliminary studies of 
the quantization of spacetimes with the symmetry of these waves have
been carried out by Neville\cite{NEV1,NEV} and Borissov\cite{BOR94}, 
mainly in the context of the Ashtekar formalism for gravity. 
Here, we will start with the geometrodynamic 
formulation for general relativity, and discuss the quantization of 
pure gravitational plane waves as a particular case of spacetimes 
that possess two commuting spacelike Killing vectors.

The rest of the paper is organized as follows. The geometrical 
properties of pure gravitational plane waves are reviewed in Sec. II.
With a suitable choice of (local) coordinates, we cast the metric of 
this family of solutions in a convenient (3+1) form. This metric can 
be written in terms of two arbitrary functions, and corresponds to 
spacetimes with two commuting Killing vector fields that are subject
to certain conditions. In Sec. III, we consider the Hamiltonian 
formulation for spacetimes that admit two commuting spacelike Killing
vectors and perform a partial gauge fixing. The reduction of the 
system is completed in Sec. IV, where we impose gauge-fixing and 
symmetry conditions that remove all non-physical degrees of
freedom and guarantee that the classical solutions are precisely 
pure gravitational plane waves. We determine the reduced phase space
of the model obtained in this way and its symplectic form. 
It is also shown that, in an appropriate set of coordinates, 
the dynamical evolution of this reduced model is trivial, so
that all relevant information about the system is in fact contained 
in the symplectic structure. In Sec. V, we particularize our study 
to the family of linearly polarized plane waves in vacuum 
gravity. We prove that the symplectic form on the reduced phase 
space of this model can be regarded as that corresponding to 
an infinite set of annihilation and creation like variables. We then 
proceed to quantize the model by introducing a Fock representation. 
Finally, we summarize our results and conclude in Sec. VI.

\section{Pure gravitational plane waves}
\setcounter{equation}{0}

Gravitational plane waves are solutions to Einstein field equations
that possess a five-parameter group of isometries\cite{BPR}. For 
these spacetimes, the metric can always be written in the 
form\cite{MAC}:
\begin{equation}\label{harmonicds2}
ds^2=-dU dV +  H_{ab}(U) X^a X^b dU^2 + \sum_a(dX^a)^2,
\end{equation}
where we have employed the index 
notation $a,b=1,2$. The coordinates $\{U,V,X^1,X^2\}$ are called 
harmonic coordinates, and run over the whole real axis. It is easy 
to check that $\partial_V$ corresponds to a covariantly constant null
vector field, so that metrics (\ref{harmonicds2}) are a special 
class of pp-waves\cite{MAC}. On the other hand, the only 
non-vanishing component of the Ricci tensor for these metrics is 
$R_{UU}= -(H_{11}+H_{22})$. Solutions to Einstein equations in 
vacuum are hence given by all symmetric matrices $H_{ab}$ with 
vanishing trace. In this case, the plane wave is said to be purely 
gravitational. 

Although the use of harmonic coordinates for plane waves has clear 
advantages, because they allow to cover the totality of the spacetime
with a single chart and express the Ricci tensor in terms of a simple
metric function, the spacetime symmetries become much more 
apparent in the so-called group coordinates. In such coordinates, 
the metric reads\cite{MAC,VER}:
\begin{equation} \label{groupds2}
ds^2=-dU d\bar{V} +  h_{ab}(U) dx^a dx^b.
\end{equation}
Here, $\bar{V}$, $x^1$, and $x^2$ run over the entire real line, and
the coordinate $U$ has in general a restricted range, as we will see 
below. Obviously, $\partial_{x^1}$ and $\partial_{x^2}$ are two 
commuting spacelike Killing vector fields for the above metrics, and
the group orbits spanned by them are everywhere orthogonal 
to the surface with coordinates $U$ and $\bar{V}$. It is 
convenient to write the two-metric $h_{ab}$ of these group orbits as
\begin{eqnarray} \label{hab}
h_{11}&=&e^{z-y/2},\hspace*{.4cm}h_{12}=-v e^{z-y/2},\nonumber
\\ h_{22}&=&(v^2+e^y)e^{z-y/2}.
\end{eqnarray}
Hence, the condition that $h_{ab}$ is positive definite is 
automatically satisfied provided that the functions $v$, $y$, and 
$z$ are real. Note that the determinant of $h_{ab}$ is equal to 
$e^{2z}$. Thus, from our discussion in the Introduction, we have 
$W=e^z$ and, since this function depends only on the coordinate $U$, 
its gradient is null with respect to the metric (\ref{groupds2}), as 
we had anticipated. Finally, the functions $v$ and $y$ describe, 
respectively, the diagonal and non-diagonal degrees of freedom of 
the metric $e^{-z}h_{ab}$, which has unit determinant. 

In group coordinates, the only non-trivial Einstein equation in 
vacuum is
\begin{equation}
R_{UU}= -{1\over 2} h^{ab} \frac{d^2h_{ab}}{dU^2} -{1\over 4}
\frac{dh^{ab}}{dU}\frac{dh_{ab}}{dU}=0, 
\end{equation}
with $h^{ab}$ being the inverse of the two-metric $h_{ab}$.
Substituting Eq. (\ref{hab}) in this expression, we get 
\begin{equation}\label{con}
\frac{d^2(e^{z/2})}{dU^2}=-\frac{e^{z/2}}{16}\left[\left(\frac{dy}
{dU}\right)^2+4\left(\frac{dv}{dU}\right)^2e^{-y}\right].
\end{equation}
Therefore, $W^{1/2}=e^{z/2}$ must be a convex function of the 
coordinate $U$. Since this function is non-negative, it must then
generally vanish at some point. The metric becomes degenerate at that
point and a coordinate singularity appears. As a consequence, the 
spacetime for pure gravitational plane waves cannot be described by 
a unique chart in group coordinates\cite{VER}.

Actually, the fact that $W^{1/2}$ is a non-negative convex function 
implies that it must vanish at either one or two points in the 
allowed (simply connected) interval of variation for the coordinate
$U$. The only exception is pure Minkowski spacetime, where $v$, 
$y$, and $z$ are constant. When $W^{1/2}$ has two zeros, it
increases from the first of them until a maximum is reached. This
maximum can always be made equal to the unity by means of a 
suitable rescaling of the real coordinates $x^a$. On the other hand, 
when $W^{1/2}$ has a single zero, it can be chosen as a strictly 
increasing function of $U$ by performing a reversal of the null 
coordinates $U$ and $V$, if necessary. In this case, $W^{1/2}$ will 
either tend to infinity or to a finite value in the limit of large 
positive $U$'s. If the limit is finite, an appropriate rescaling of 
$x^a$ sets it equal to the unity. Hence, we see that in all cases 
there exists a region of the spacetime where the function $W^{1/2}$ 
increases with $U$ and ranges in the interval $(0,1)$. In this 
region, $z$ is strictly increasing from minus infinity to zero.
In addition, although Minkowski spacetime has apparently been 
excluded from our discussion, it turns out that the above 
considerations apply as well to flat spacetime, because the solution
to Eq. (\ref{con}) in which $v$ and $y$ are constant and $W^{1/2}$ 
is linear in $U$ can be cast in Minkowskian form by a (local) 
coordinate transformation\cite{BPR}.

In the region where $z$ increases in $(-\infty,0)$,
we can introduce a change of coordinates 
\begin{equation} \label{chu}
dU=e^{2w-z+y/2}du
\end{equation}
so that the function $z$ gets fixed in terms of our new coordinate 
$u$. For instance, we can make
\begin{equation} \label{z0}
z(U[u])= -e^{-u}\equiv z_0(u).
\end{equation}
In Eq. (\ref{chu}), $w$ is an unknown function of $u$. This 
coordinate is assumed to run over the entire real axis. Our change 
of coordinates is then well-defined and invertible, because $z_0$ is 
strictly increasing for $u\in I\!\!\!\,R$ and has the range 
$I\!\!\!\,R^-$. With this change, the coordinate singularity at 
$e^z=0$ has been driven to minus infinity. Notice also that our 
ignorance about the original function $z(U)$ is encoded in $w$. 
However, the vacuum Einstein equation (\ref{con}) becomes now a 
first-order differential equation in $w$ that can be solved 
explicitly. The solution is
\begin{eqnarray} \label{w0}
w&=&\frac{\ln{(z_0^{\prime})}}{2}+\frac{3z_0}{4}-\frac{y}{4}+ 
\int^u_{u_0}\frac{1}{16z^{\prime}_0}\left[(y^{\prime})^2+
4(v^{\prime})^2 e^{-y}\right]\nonumber\\ &&\equiv w_0,
\end{eqnarray}
where $u_0$ is any fixed real constant and the prime stands for
derivative with respect to $u$. The constant of integration that 
should appear in this formula has been absorbed by a scale 
transformation in the coordinate $\bar{V}$. 

If we finally define a time via $\bar{V}=2t-u$, we arrive at 
the following line element for pure gravitational plane waves:
\begin{equation}\label{grouptds2}
\!ds^2\!=e^{2w_0-z_0+y/2}[-dt^2\!+\!(du-dt)^2]+h_{ab}dx^adx^b\!,
\end{equation}
with $h_{ab}$ the metric obtained from Eq. (\ref{hab}) by setting 
$z=z_0$. All coordinates in this expression are real and have 
unrestricted ranges. Recalling relations (\ref{z0}) and (\ref{w0}), 
we see that this family of vacuum solutions have only two degrees of
freedom, which are given by the functions $v$ and $y$ of the 
coordinate $u$.

These are the spacetimes that we will study in the rest of this 
paper. As we have shown, the above metric describes the most general 
pure gravitational plane wave solution, with the only caveat that it
does not correspond to the whole of the spacetime that can be covered
with harmonic coordinates. The coordinate system that we have adopted
is in fact quite similar to that employed in Ref.\cite{BPR}, the 
main difference being that we have chosen $z_0=-e^{-u}$ and allowed 
$u$ to take all real values, while $z_0$ was made equal to 
$2\ln{|u|}$ in Ref.\cite{BPR}. In this latter case, $u$ has a 
restricted range [e.g., $u\in(0,1)$ for $z_0\in I\!\!\!\,R^-$].

We can regard Eq. (\ref{grouptds2}) as providing the standard $3+1$
decomposition of the metric for spacetimes with two orthogonal 
surfaces, but with the lapse and (non-vanishing component of) 
the shift vector particularized to take the values 
$N=e^{w_0-z_0/2+y/4}$ and $N^u=-1$. In addition, the spacetimes 
considered admit two commuting spacelike Killing vector fields,
$\partial_{x^1}$ and $\partial_{x^2}$, and satisfy another 
condition, namely, that the metric functions are independent not 
only of $x^a$, but also of the time coordinate $t$. In the 
Hamiltonian formalism for general relativity, there is a simple way 
of imposing the time independence of the metrics (\ref{grouptds2}) 
on classical solutions, once the existence of the two spacelike 
Killing vectors has been assumed. We first note that $z_0$ is 
a fixed function of $u$, and that $w_0$ is determined by the vacuum 
Einstein equations to be given by Eq. (\ref{w0}) in terms of $v$ 
and $y$. It hence suffices to ensure that the time derivatives 
$\dot{v}$ and $\dot{y}$ vanish on classical solutions. Employing then
the definition of the extrinsic curvature $K_{ij}$\cite{WA}, with 
${x^i}\equiv\{x^1,x^2,u\}$, it is not difficult to check that, for 
the metrics under discussion, the requirements $\dot{v}=\dot{y}=0$ 
on classical solutions are equivalent to the relations
\begin{eqnarray}\label{K12}
4h^{1/2}e^{-2z_0+y/2}K_{11}&=&2z_0^{\prime}-y^{\prime},\nonumber \\ 
4h^{1/2}e^{-2z_0+y/2}K_{12}&=&vy^{\prime}-2vz_0^{\prime}-2v^{\prime},
\end{eqnarray}
where $h=e^{2w_0+z_0+y/2}$ is the determinant of the induced metric 
$h_{ij}$. Moreover, these relations turn out to imply the vanishing 
of $\dot{v}$ and $\dot{y}$ even when the values of $w_0$, the lapse 
function, and the shift component $N^u$ are changed in the metrics 
(\ref{grouptds2}), provided that $N^u$ does not depend on the 
coordinates $x^a$ and the quotient $N/N^u=-h^{1/2}e^{-z_0}$ is not 
modified.

\section{Spacetimes with two Killing vector fields}
\setcounter{equation}{0}

In this section, we will discuss the Hamiltonian formulation for 
spacetimes that possess two commuting spacelike Killing vectors, 
paying a special attention to the case of pure gravitational 
plane waves. 

We will only consider spacetimes that admit a global foliation, 
so that the metric can be written in the (3+1) form:
\begin{equation} 
ds^2=-N^2dt^2+h_{ij}(dx^i+N^idt)(dx^j+N^jdt).
\end{equation}
For the coordinates of the sections of constant time, we will employ 
the notation $\{x^i\}\equiv\{x^1,x^2,u\}$. In addition, we will 
assume that the coordinates $t$ and $u$ run over the real line, and 
impose that $\partial_{x^a}$ ($a=1,2$) are the two spacelike Killing 
vector fields. As a consequence, the metric must be independent of 
the coordinates $x^a$. This implies that the area
${\cal S}=\int dx^1dx^2$ appears as a constant global factor in the 
gravitational Hilbert-Einstein action and in the symplectic form for 
geometrodynamics. We can always absorb this area in 
Newton's constant. In order to do this, renormalization is required 
when the area ${\cal S}$ is infinite\cite{NEV1} (this case can be 
considered as the limit ${\cal S}\rightarrow\infty$ of the compact 
case). For convenience, we further set the effective Newton's 
constant obtained in this way equal to $1/8\pi$. 

The fundamental Poisson brackets and momenta ca\-no\-nically
conjugate to $h_{ij}$ are then given by
\begin{eqnarray}
&&\{h_{ij}(u),\Pi^{kl}(\bar{u})\}=\delta^{(k}_{\;i}\delta^{l)}_{j}
\delta(u-\bar{u}),\\ \label{mok}
&&\Pi^{ij}=\frac{1}{2}h^{1/2}(h^{ik}h^{jl}-h^{ij}h^{kl})K_{kl},
\end{eqnarray}
where $\delta^i_j$ is the Kronecker delta, $\delta(u)$ is the Dirac 
delta on the real line, the indices in parenthesis are symmetrized, 
and $h^{ij}$ is the inverse of the induced metric. Thus, the 
symplectic form on phase space is
\begin{equation} 
\Omega=\int du\, {\bf d} \Pi^{ij}\wedge {\bf d} h_{ij},
\end{equation} 
with the symbols ${\bf d}$ and $\wedge$ denoting the exterior 
derivative and product, respectively.

The gravitational system has two kinds of constraints: 
the Hamiltonian constraint ${\cal H}$ and the momentum
constraints ${\cal H}_i$, which generate spatial diffeomorphisms. 
Instead of dealing with ${\cal H}$, we will consider its densitized 
version, $\tilde{\cal H}\equiv h^{1/2}{\cal H}$. In terms of 
phase-space variables, these constraints take the expressions
\begin{eqnarray} 
&&\tilde{\cal H}\equiv -\frac{h}{2}\,^{(3)}R+(h_{ik}h_{jl}+h_{il}
h_{jk}-h_{ij}h_{kl})\Pi^{ij}\Pi^{kl}=0,\nonumber\\
&&{\cal H}_i\equiv -2h_{ij} D_k\Pi^{kj}=0.
\end{eqnarray}
Here, $D_i$ is the covariant derivative compatible with the induced 
metric and $^{(3)}R$ its curvature scalar. The time 
derivative of any function on phase space is then
\begin{equation}
\dot{f}=\partial_{t}f+\{f,\int du\,(N_{_{_{\!\!\!\!\!\!\sim}}\;}
\tilde{\cal H}+N^i{\cal H}_i)\},
\end{equation}
where $\partial_t$ is the partial derivative with respect to the
explicit time dependence and $N_{_{_{\!\!\!\!\!\!\sim}}\;}=h^{-1/2}N$
is the densitized lapse function. 

Using that the metric does not depend on the coordinates $x^a$, it 
is not difficult to check that the momentum constraints ${\cal H}_a$
can be rewritten as
\begin{equation} 
{\cal H}_a=-2\left(h_{ai}\Pi^{iu}\right)^{\prime}.
\end{equation}
For the globally hyperbolic spacetimes with two commuting spacelike 
Killing vectors that we are studying, it is then possible to fix the 
gauge freedom associated with these constraints by demanding that, 
on the sections of constant time, the $u$-line is orthogonal to the 
group orbits spanned by $\partial_{x^a}$. This requirement can be 
rephrased in the form
\begin{equation} \label{gfc1}
h_{au}=0,\hspace*{.4cm} a=1,2. 
\end{equation}

To see that the above conditions lead to a well-posed gauge fixing,
we must prove that, in general, the gauge orbits generated by the 
constraints ${\cal H}_a$ intersect transversely the surface that is 
defined on phase space by conditions (\ref{gfc1}) and the 
constraints of the system. This is equivalent to prove that the 
gauge-fixing conditions have non-vanishing Poisson brackets with 
${\cal H}_a$. A straightforward 
calculation shows
\begin{equation} 
\{h_{au},\int du\, N^b {\cal H}_b\}=h_{ab}(N^b)^{\prime}.
\end{equation}
Taking into account that $h_{ab}$ is positive definite, we 
conclude that these brackets are generally different from zero 
(it suffices to take $(N^a)^{\prime}\neq 0$). 

We must also check that our gauge fixing is compatible with 
dynamics, in the sense that there exists a choice 
for the components $N^a$ of the shift vector (which are the Lagrange 
multipliers of ${\cal H}_a$) so that, on the constraint surface, the 
gauge conditions (\ref{gfc1}) are preserved by the dynamical 
evolution. In other words, modulo constraints and gauge-fixing 
conditions, $\dot{h}_{au}$ must vanish for a particular choice of 
$N^a$. Recalling the independence of the metric on $x^a$, one can 
see that
\begin{equation}
\dot{h}_{au}=h_{ab}(N^b)^{\prime}+4N_{_{_{\!\!\!\!\!\!\sim}}\;}
h_{ab}h_{uu}\Pi^{bu},
\end{equation}
where use of the gauge conditions has been made after calculating 
Poisson brackets. On the other hand, the solution to ${\cal H}_a=0$ 
is given by
\begin{equation} 
\Pi^{au}=h^{ab}f_b(t),
\end{equation}
with $f_a$ being two arbitrary functions of $t$ that must be real, 
because $\Pi^{au}\in I\!\!\!\,R$. Thus, our gauge fixing 
is in fact preserved by the evolution provided that
\begin{equation}
(N^a)^{\prime}=-4N_{_{_{\!\!\!\!\!\!\sim}}\;}h_{uu}h^{ab}f_b(t).
\end{equation}

Let us now integrate both terms in the above equality over 
$u$, and contract the result with $f_a(t)$. If we 
assume that the components $N^a$ of the shift vanish in the limit of 
large $u$'s, we get
\begin{equation} \label{vfs} 
\int_{I\!\!\!\,R}du\,
N_{_{_{\!\!\!\!\!\!\sim}}\;}h_{uu}h^{ab}f_af_b=0.
\end{equation}
It is worth commenting that, together with the conditions $h_{au}=0$,
the requirement $\lim_{u\rightarrow\pm\infty}N^a=0$ can be viewed as
ensuring that the group orbits spanned by $\partial_{x^a}$ are 
asymptotically orthogonal to the surfaces with coordinates $t$ and 
$u$. This requirement should hence be satisfied at least in 
spacetimes where those surfaces are everywhere orthogonal. Returning
to Eq. (\ref{vfs}), we note that, since the integrand in that 
expression is real, it has to vanish at least at one point if the 
integral is equal to zero. Taking into account that the two-metric 
$N_{_{_{\!\!\!\!\!\!\sim}}\;}h_{uu}h_{ab}$ is positive definite, it 
follows that $f_a=0$ at that point. The functions $f_a$ must then be
identically zero, because they are independent of $u$. Therefore, 
one obtains that, when $N^a$ vanishes for large absolute values of 
$u$, the solutions to the constraints ${\cal H}_a=0$ and consistency 
conditions $\dot{h}_{au}=0$ are
\begin{equation} 
\Pi^{au}=0,\hspace*{.4cm} N^a=0.
\end{equation}
The vanishing of $h_{au}$ and $N^a$ implies that the spacetime 
metric is block-diagonal, so that the assumption 
$\lim_{u\rightarrow\pm\infty}N^a=0$ employed in our gauge fixing 
turns out to guarantee that the surface with coordinates $x^a$ is 
actually orthogonal to that described by $t$ and $u$. 

After this partial gauge fixing, the canonically conjugate pairs 
$(h_{au},\Pi^{au})$ and the momentum constraints ${\cal H}_a$ are 
eliminated from the system. The remaining geo\-me\-tro\-dynamic 
momenta and components of the induced metric provide a canonical set
of variables on the phase space of the gauge-fixed model. The line 
element of this reduced model has the form
\begin{equation} 
ds^2=-N^2dt^2+h_{uu}(du+N^udt)^2+h_{ab}dx^adx^b,
\end{equation}
where the metric functions depend only on the coordinates $t$ and 
$u$.

At this stage, it is convenient to carry out a change of metric 
variables from $h_{uu}$ and $h_{ab}$ to a set of functions
$\{q^\alpha\}\equiv\{v,w,y,z\}$ that is analogous to that employed 
to describe the family of metrics (\ref{grouptds2}). In this way, 
our discussion will be straightforwardly applicable to the analysis 
of pure gravitational plane waves. The change is such that the 
line element adopts the expression
\begin{eqnarray}\label{dsgf}
ds^2=e^{2w-z+y/2}[-e^{2z}N_{_{_{\!\!\!\!\!\!\sim}}\;}^2
dt^2+(du+N^udt)^2]+  \nonumber
\\ e^{z-y/2}[(dx^1)^2-2vdx^1dx^2+(v^2+e^y)(dx^2)^2].
\end{eqnarray}
Here, all metric functions are allowed to depend on the coordinates 
$u$ and $t$, and we have densitized the lapse using that
\begin{equation}
h=e^{2w+z+y/2}.
\end{equation}
Similarly to the situation found for plane waves, the condition that 
the induced metric is positive definite is now easily imposed: it 
suffices to demand that the functions $q^{\alpha}$ are real. On the 
other hand, since the change of variables performed is just a point 
transformation, it is possible to find a set of momenta 
$\{p_{\alpha}\}$ that are canonically conjugate to our new metric 
variables. These momenta are 
\begin{equation} \label{palp}
p_{\alpha}=\Pi^{uu}\frac{\partial h_{uu}}{\partial q^{\alpha}\;}
+\Pi^{ab}\frac{\partial h_{ab}}{\partial q^{\alpha}\;},
\end{equation}
where $h_{uu}$ and $h_{ab}$ are regarded as functions of 
$q^{\alpha}$. Using relation (\ref{mok}), we then obtain that, in 
terms of the extrinsic curvature and our metric variables,
\begin{eqnarray}\label{pks} 
&&p_v=-h^{1/2}e^{-z-y/2}(vK_{11}+K_{12}), \nonumber\\
&&p_w=-h^{1/2}e^{-z-y/2}[(v^2+e^y)K_{11}+2vK_{12}+K_{22}],\nonumber\\
&&p_y=-\frac{1}{2}h^{1/2}e^{-z+y/2}K_{11}, \nonumber\\
&&p_z=-h^{1/2}e^{-2w+z-y/2}K_{uu}.
\end{eqnarray}

Finally, the gravitational model considered is still subject to two 
constraints, which are given by the restriction of ${\cal H}_u$ and
$\tilde{\cal H}$ to the case in which the metric is independent of 
the coordinates $x^a$ and $h_{au}=\Pi^{au}=0$. After some trivial 
calculations, one can show that the momentum constraint ${\cal H}_u$ 
takes the form
\begin{equation} 
{\cal H}_u=-p_w^{\prime}+p_w w^{\prime}+
p_v v^{\prime}+p_y y^{\prime}+p_z z^{\prime},
\end{equation}
and that the densitized Hamiltonian constraint has the expression
\begin{eqnarray}
\!\tilde{\cal H}&=&\!\frac{e^{2z}}{16}\left[(y^{\prime})^2+
4(v^{\prime})^2e^{-y}-4z^{\prime}(4w^{\prime}+y^{\prime}-5
z^{\prime})+16z^{\prime\prime}\right]\nonumber\\
&&+p_v^2e^y-2p_wp_y-p_wp_z+4p_y^2.  
\end{eqnarray}

\section{Gauge fixing and symmetry reduction}
\setcounter{equation}{0}

We will now apply the analysis of the previous section to the study 
of pure gravitational plane waves. We will complete the gauge fixing 
of the system and introduce symmetry conditions that ensure that all 
classical solutions are plane waves in vacuum, so that the reduced 
model that is obtained really describes the family of metrics 
(\ref{grouptds2}).

Let us first remove the gauge freedom associated with the constraint 
${\cal H}_u$, which generates diffeomorphisms in the coordinate $u$. 
A way to fix this gauge is to provide a one-to-one correspondence 
between the coordinate $u$ and a metric variable. This can be done in
the case of the plane wave metrics that we are considering, because 
the variable $z$ can then be chosen as a strictly increasing 
function of $u\in\!\!\!\,R$, namely, the function $z_0$ defined in 
Eq. (\ref{z0}). In order to fix the gauge freedom, we then introduce
the condition 
\begin{equation}\label{chiu} 
\chi_u\equiv z-z_0=z+e^{-u}=0.
\end{equation}
Since $z_0^{\prime} \neq 0$, the constraint ${\cal H}_u=0$ can be
easily solved to find the momentum canonically conjugate to $z$:
\begin{equation}
p_z=p_z^0\equiv \frac{1}{z_0^{\prime}} (p_w^{\prime}-p_w
w^{\prime}-p_v v^{\prime}-p_y y^{\prime}).
\end{equation}
Therefore, if our gauge fixing is admissible, we can eliminate the 
canonical pair $(z,p_z)$ from the system.

Actually, condition (\ref{chiu}) leads to a well-posed gauge fixing,
because (modulo that condition)
\begin{equation} 
\{\chi_u,\int du\, N^u {\cal H}_u\}=z_0^{\prime}N^u.
\end{equation}
Obviously, this Poisson bracket differs from zero when $N^u\neq 0$, 
because $z^{\prime}_0$ does not vanish at any point. On the other 
hand, if our gauge condition is compatible with the dynamical 
evolution, there must exist a choice for $N^u$ (i.e., the Lagrange 
multiplier of ${\cal H}_u$) such that
\begin{eqnarray} 
\dot{\chi}_u&=&\{\chi_u,\int du\,(N_{_{_{\!\!\!\!\!\!\sim}}\;}
\tilde{\cal H}+N^u{\cal H}_u)\}\nonumber\\
&=&-p_wN_{_{_{\!\!\!\!\!\!\sim}}\;}+z_0^{\prime}N^u=0. 
\end{eqnarray}
In the second line of this equation, we have substituted $z=z_0$.
The condition $\chi_u=0$ is thus preserved by the evolution if
\begin{equation} \label{Nu}
N^u=\frac{p_wN_{_{_{\!\!\!\!\!\!\sim}}\;}}{z_0^{\prime}}.
\end{equation}
Note that the component $N^u$ of the shift given by this formula is 
always well-defined, because $z_0$ never vanishes. We hence conclude
that the gauge fixing introduced is acceptable. It is worth remarking
that, to arrive at this result, the only property of the function 
$z_0$ that has been used is that its derivative with respect to
$u$ is everywhere finite and different from zero. In this sense, the
explicit form of this function, as well as its range, are irrelevant.

After this gauge fixing, we obtain a reduced model whose phase 
space is described by the canonical set of variables 
$\{v,w,y,p_v,p_w,p_y\}$. The symplectic form is 
\begin{equation}\label{sfr}
\Omega=\int du\,({\bf d}p_v\wedge {\bf d}v+{\bf d}p_w\wedge{\bf d}w+
{\bf d}p_y\wedge {\bf d}y),
\end{equation}
and the system is subject to the densitized Hamiltonian constraint
\begin{equation} 
\tilde{\cal H}_r\equiv \tilde{\cal H}(z=z_0,p_z=p_z^0)=0.
\end{equation} 

In order to complete the reduction and attain a midisuperspace model
for pure gravitational plane waves, we still need to introduce two 
types of conditions. Firstly, we have to impose that all classical 
solutions of the system are independent of the time coordinate, so 
that these solutions correspond to the family of plane wave metrics 
(\ref{grouptds2}). Secondly, we have to find a condition that 
results in removing the gauge freedom associated with the densitized 
Hamiltonian constraint. Both tasks can be achieved in the following 
way.  

We have shown at the end of Sec. II that, for metrics of the form 
(\ref{dsgf}) with $z=z_0$ and 
$N^u=-e^{z_0}N_{_{_{\!\!\!\!\!\!\sim}}\;}$, the time 
independence of the metric functions $v$ and $y$ on classical 
solutions is ensured by relations (\ref{K12}). Using 
Eqs. (\ref{pks}), those relations can be expressed as the conditions
\begin{eqnarray} 
\chi_1&\equiv&p_y-\frac{e^{z_0}}{8}(y^{\prime}-2z^{\prime}_0)=0, 
\nonumber\\
\label{X2}\chi_2&\equiv& 2 e^{y-z_0}p_v-v^{\prime}=0.
\end{eqnarray}
Taking into account identity (\ref{Nu}), we see 
that these conditions guarantee that $v$ and $y$ do not depend 
classically on time provided that 
\begin{equation}\label{X0}
\chi_0\equiv p_w+z_0^{\prime}e^{z_0}=0.
\end{equation}
When this is the case, it is clear that conditions (\ref{X2})
will also imply the time independence of the momenta 
$p_v$ and $p_y$. In addition, the momentum $p_w$ will not depend on 
time if the requirement $\chi_0=0$ is satisfied. The only phase-space
variable that might then be time dependent on classical solutions is 
$w$. This possible time dependence can nevertheless be eliminated by
demanding that 
\begin{equation} 
\chi_3\equiv w-w_0=0,
\end{equation}
where $w_0$ is the function defined in Eq. (\ref{w0}). In this way,
the function $w$ is fixed to take the same expression in terms of
$v$ and $y$ as it adopts in pure gravitational plane waves. On the 
other hand, the constraint of the model can be rewritten after some 
manipulations as
\begin{eqnarray} \label{Hr}
\tilde{\cal H}_r&=&(2y^{\prime}\chi_1+e^{z_0-y}v^{\prime}
\chi_2+2w^{\prime}\chi_0-2\chi_0^{\prime})\frac{\chi_0}
{2z_0^{\prime}}+4\chi_1^2\nonumber \\
&&+ \frac{e^{2z_0-y}}{4}\chi_2^2 -2(\chi_1+e^{z_0}
\chi_3^{\prime})\chi_0+e^{z_0}\chi_0^{\prime}.
\end{eqnarray}
So, once the symmetry conditions $\chi_1=\chi_2=0$ are imposed, the 
requirement $\chi_0=0$ is just a solution to $\tilde{\cal H}_r=0$. 
Therefore, the surface defined on phase space by $\chi_I=0$ 
($I=0,...,3$) is simply a section of the constraint surface. In the 
following, we will show that the reduction to this section is 
consistent and that, in the reduction process, the condition 
$\chi_3=0$ allows us to fix the gauge freedom associated with 
$\tilde{\cal H}_r$. We will employ the symbol $\approx$ for weak 
identities, namely, identities that are satisfied modulo $\chi_I=0$. 
Besides, we will denote the set 
$\{\tilde{\cal H}_r,\chi_1,\chi_2,\chi_3\}$ by $\{\chi_{\cal A}\}$,
with ${\cal A}=0,...,3$.

To prove that the reduction is admissible, we first have to show 
that the conditions and constraint $\chi_{\cal A}=0$ constitute a 
second-class system on the section of the constraint surface 
that we are analyzing. Let us define
\begin{equation} 
\chi(g)=\int du \sum_{\cal A} g_{\cal A} \chi_{\cal A}, 
\end{equation}
where $\{g_{\cal A}\}$ is any set of $C^{\infty}_0$ functions of the 
coordinate $u$, i.e., functions of $u$ that are infinitely 
differentiable and have compact support. Clearly, imposing 
$\chi_{\cal A}=0$ is equivalent to demand that $\chi(g)$ vanishes 
for all choices of the functions $g_{\cal A}$. The set 
$\{\chi_{\cal A}\}$ is then second-class provided that no combination
of the form $\chi(g)$ (other than the zero constant) commutes weakly 
under Poisson brackets with all of the $\chi_{\cal A}$'s. Using 
expression (\ref{Hr}) for the constraint, and taking into account 
that all terms in that expression are quadratic in $\chi_I$ except 
the last one, we obtain
\begin{equation} 
\{\tilde{\cal H}_r,\chi(g)\}\approx -e^{z_0}g_3^{\prime}.
\end{equation}
Since $e^{z_0}\neq 0$ and the function $g_3$ is $C_0^{\infty}$, it 
follows that $\chi(g)$ commutes weakly with $\tilde{\cal H}_r$ if 
and only if $g_3$ vanishes. Let now $\bar{\chi}(g)$ be $\chi(g)$ for 
$g_3=0$. After some calculations, one gets
\begin{equation} 
\sum_{{\cal A}=1}^2 g_{\cal A}\{\chi_{\cal A},\bar{\chi}(g)\}\approx 
-\frac{1}{8}(e^{z_0}g_1^2+16e^{y-z_0}g_2^2)^{\prime}.
\end{equation}
Given that the term in parenthesis is a sum of non-negative 
functions and that $g_1,g_2\in C_0^{\infty}$, the vanishing of the 
Poisson brackets of $\bar{\chi}(g)$ with $\chi_1$ and $\chi_2$ turns 
out to imply that $g_1$ and $g_2$ must be the zero function. On the 
other hand, it is not difficult to check that
\begin{equation} 
\{\chi_3,\hat{\chi}(g)\}\approx -(e^{z_0}g_0)^{\prime},
\end{equation}
where $\hat{\chi}(g)$ denotes the restriction of $\chi(g)$ to the 
case in which $g_0$ is the only function in $\{g_{\cal A}\}$ that 
differs from zero. Recalling that $g_0\in C^{\infty}_0$, we see that
the above Poisson bracket vanishes weakly provided that $g_0=0$. 
Therefore, $\chi(g)$ must be exactly equal to zero in order to
commute weakly with the set $\{\chi_{\cal A}\}$, as we wanted to 
prove.

Let us now show that the section of the constraint surface determined
by $\chi_I=0$ ($I=0,...,3$) is preserved by the dynamical evolution 
for a suitable choice of the densitized lapse. From expression 
(\ref{Hr}) and the fact that the variables 
$\{\rho^{\alpha}\}\equiv\{v,y,p_v,p_w,p_y\}$ commute with $\chi_0$, 
we get
\begin{equation} 
\dot{\rho}^{\alpha}=\{\rho^{\alpha},\int du\,
N_{_{_{\!\!\!\!\!\!\sim}}\;}\tilde{\cal H}_r\}\approx 0.
\end{equation}
As a consequence, the variables $\rho^{\alpha}$ turn out to be time 
independent on classical solutions. To deduce this result, one 
actually needs not demand that $\chi_3$ vanishes. Moreover, one can 
then straightforwardly see that the symmetry conditions 
$\chi_1=\chi_2=0$ 
are preserved in time, and that, when these conditions are imposed, 
the solution to the constraint given by $\chi_0=0$ is invariant under
the evolution, regardless of the value taken by the densitized lapse. 
On the other hand, employing again the expression for the 
constraint and that $\{\chi_3,\chi_0\}=\{w,\chi_0\}$, we obtain
\begin{equation} 
\dot{\chi}_3\approx -(e^{z_0}N_{_{_{\!\!\!\!\!\!\sim}}\;})^{\prime}
\approx \dot{w}.
\end{equation}
So, compatibility of the condition $\chi_3=0$ with the dynamical 
evolution ensures that the variable $w$ does not depend classically 
on time, and implies
\begin{equation} 
N_{_{_{\!\!\!\!\!\!\sim}}\;}=F(t)e^{-z_0}.
\end{equation}
The arbitrary function $F$ that appears in this expression can in 
fact be absorbed by means of the time redefinition 
$\tau=\int^t F(\bar{t})d\bar{t}$. In this way, we arrive at 
$N_{_{_{\!\!\!\!\!\!\sim}}\;}=e^{-z_0}$. From Eqs. (\ref{Nu}) and 
(\ref{X0}), the only non-vanishing component of the shift vector is 
then given by $N^u=-1$. We hence see that the reduction proposed for
the system is acceptable and that the condition $\chi_3=0$ allows us
to fix the value of the densitized lapse function. Therefore, this 
condition removes the gauge freedom associated with the constraint 
$\tilde{\cal H}_r$. In addition, we note that, in agreement with our
previous discussion, all phase-space variables are indeed independent
of the time coordinate on classical solutions after reducing the 
system to the section of the constraint surface defined by 
$\chi_I=0$. 

In this reduction, the pair $(w,p_w)$ is eliminated via the 
gauge-fixing condition $\chi_3=0$ and the solution to the constraint 
$\chi_0=0$, while the symmetry conditions $\chi_1=\chi_2=0$ 
determine the momenta $p_v$ and $p_y$ in terms of the variables $v$ 
and $y$. The reduced model that is attained is free of constraints 
and possesses only two degrees of freedom, which are described by 
the metric variables $v$ and $y$. The line element for this reduced 
model can be obtained from Eq. (\ref{dsgf}) by substituting $z=z_0$, 
$w=w_0$, and the values obtained with our gauge fixing for $N^u$ and
the densitized lapse function. This line element coincides formally 
with that given in Eq. (\ref{grouptds2}) for pure gravitational 
plane waves. The only difference is that the functions $v$ and $y$ 
of the coordinate $u$ may now depend also on time. However, this 
time dependence is ruled out on classical solutions, as we have 
shown above. The metric may depend on time only when non-classical 
trajectories are allowed, e.g., in a quantum theory. Besides, it is 
clear that the set of classical solutions for our reduced model is 
precisely the family of gravitational plane waves in vacuum 
considered in Sec. II. As a consequence, our model describes in 
fact that family of plane waves. 

Obviously, the dynamics of this reduced model is trivial, because 
the variables $v$ and $y$ remain constant in time under the classical 
evolution. The reduced Hamiltonian, which generates the 
time evolution via Poisson brackets, must hence vanish. Of course, 
the same conclusion is reached by considering the reduced action for
the system. This action can be obtained from the Hilbert-Einstein 
action in the following manner. We adopt the (3+1) decomposition of 
the metric explained in Sec. III, impose that the metric is 
independent of $x^a$ ($a=1,2$), and absorb the area that corresponds 
to these coordinates in Newton's constant, setting the resulting 
effective constant equal to $1/8\pi$. Since all gravitational 
constraints are eliminated in the reduction process, we can make them
equal to zero in the Hamiltonian form of the action. The reduced 
action $S_r$ that we are seeking can then be computed from the 
Lagrangian density $\Pi^{ij}\dot{h}_{ij}$. Substituting $\Pi^{au}=0$,
relations (\ref{palp}), and the values taken in our model by the 
variables $(w,z,p_v,p_w,p_y,p_z)$, we get
\begin{equation} \label{Sr} 
S_r=\int dt du\, \frac{e^{z_0}}{8}(y^{\prime}\dot{y}+4v^{\prime}
e^{-y}\dot{v}).
\end{equation}
In arriving at this result, we have disregarded surface terms that
are evaluated on sections of constant time, for they do not modify 
the reduced Hamiltonian (nor affect the symplectic structure). We 
see that the reduced action is linear in time derivatives. This 
implies that the reduced model has indeed a vanishing Hamiltonian 
in the set of coordinates adopted.

Given that the dynamics is trivial, all relevant information for 
quantizing the system is encoded in its symplectic structure. The 
symplectic form $\Omega_r$ on the phase space of the reduced model 
is provided by the pull-back of the form (\ref{sfr}) to the surface 
determined by the conditions $\chi_I=0$ ($I=0,...3$). Taking into 
account that $\chi_0=0$ ensures that $p_w$ is a fixed function of 
the coordinate $u$ (because so is $z_0$), we obtain
\begin{equation} \label{omegar}
\Omega_r=\int du\, \frac{e^{z_0}}{8}[{\bf d}y^{\prime}\wedge 
{\bf d}y+4{\bf d}(v^{\prime}e^{-y})\wedge {\bf d}v].
\end{equation}
Actually, this form could have also been deduced from the reduced
action (\ref{Sr}). The quantization of the system can then be 
attained by quantizing the algebra of Poisson brackets that follows 
from this symplectic form. In principle, the algebra seems rather 
complicated. This situation contrasts with that found in other 
midisuperspace models which describe spacetimes with two commuting 
spacelike Killing vectors, like, e.g., the Gowdy model\cite{ME}, 
where the Poisson algebra obtained after (an almost complete) gauge 
fixing is simple, although the complexity of the system shows up in 
the expression for the reduced Hamiltonian. Nonetheless, it turns out
that there exists at least a class of pure gravitational 
plane waves for which this Poisson algebra becomes 
manageable, namely, the case of linearly polarized plane waves. From
now on, we will restrict our discussion to this type of waves.

\section{Linearly polarized plane waves}
\setcounter{equation}{0}

Pure gravitational plane waves for which the metric function $v$ 
vanishes are called linearly polarized\cite{MAC}. One can actually 
impose the vanishing of $v$ as an additional symmetry on the 
midisuperspace model for plane waves in source-free gravity that we 
have studied. This can be done, for instance, by including the 
condition $\chi_4\equiv v=0$ in the process of gauge fixing and 
symmetry reduction explained in the previous section. Similar 
arguments to those presented for general plane waves in vacuum 
gravity show that the set formed by $\chi_4$ and $\chi_{\cal A}$
(with ${\cal A}=0,...,3$) is second-class on the section of the 
constraint surface defined by the requirements 
$\chi_I=\chi_4=0$ ($I=0,..3$). Here, we have employed the notation 
introduced in Sec. IV. On the other hand, we have already seen 
that the metric variable $v$ is invariant under the dynamical 
evolution when the conditions $\chi_I=0$ are satisfied, so that the 
symmetry condition $\chi_4=0$ is compatible with dynamics. 
Therefore, the reduction obtained by demanding that 
$\chi_I$ and $\chi_4$ vanish is fully consistent.

The reduced model that one obtains in this manner is such that its 
phase space can be described by a single metric variable: the field 
$y$. As it happens for plane waves with two polarizations, this 
variable depends on the coordinate $u$ and, although it might in 
principle evolve in time, its time dependence is forbidden on 
classical solutions. It is easy to check that the action $\bar{S}_r$
and the symplectic form $\bar{\Omega}_r$ for this reduced model 
coincide in fact with those given in Eqs. (\ref{Sr}) and 
(\ref{omegar}) when particularized to the case $v=0$. Defining 
\begin{equation}\label{Ylp} 
Y=\frac{e^{z_0/2}y}{2\sqrt{2}},
\end{equation}
we can then write
\begin{equation} \label{rlp} 
\bar{S}_r=\int dt du\, Y^{\prime }\dot{Y},\hspace*{.4cm}
\bar{\Omega}_r=\int du\, {\bf d}Y^{\prime}\wedge {\bf d}Y,
\end{equation}
where we have used that $z_0$ is a fixed function of $u$ and 
neglected surface contributions to the action that come from
sections of constant time. Note that the reduced action $\bar{S}_r$ 
corresponds indeed to a system with zero Hamiltonian and symplectic 
structure given by $\bar{\Omega}_r$.

Like in the case of waves with two polarizations, the dynamics of the 
reduced model for linearly polarized waves is trivial in the set of 
coordinates adopted. The function $Y$ is constant in time and 
the reduced Hamiltonian vanishes. Hence, in order to quantize the 
system, we only need to quantize the algebra of Poisson brackets 
that follows from the symplectic form $\bar{\Omega}_r$. 

On the other hand, employing that the coordinate $u$ runs over the 
whole real axis, we can express the function $Y$ as the Fourier 
transform
\begin{equation}\label{ft} 
Y=\frac{1}{\sqrt{2\pi}}\int_0^{\infty}\frac{dk}{\sqrt{2k}}\,
(a_k e^{-iku}+a_k^{\star} e^{iku}).
\end{equation}
The complex functions $a_k$ and $a_k^{\star}$, with 
$k\in I\!\!\!\,R^+$, provide then a complete set of variables on the 
phase space of the reduced model. These functions of $k$ might 
depend as well on time; however, since $Y$ is time independent on 
classical solutions, they turn out to be classical constants of 
motion. Substituting the above expression for $Y$ in the symplectic 
form, we obtain 
\begin{equation} 
\Omega_r=i\int_0^{\infty}dk\, {\bf d}a_k^{\star}\wedge {\bf d}a_k.
\end{equation}
As a consequence, the only non-vanishing Poisson brackets between 
the phase-space variables $a_k$ and $a_k^{\star}$ are 
\begin{equation}\label{pbr} 
\{a_{\bar{k}},a_{k}^{\star}\}=-i\delta(\bar{k}-k).
\end{equation}
In addition, the fact that the function $Y$ is real implies 
that $a_{k}^{\star}$ must be the complex conjugate of $a_k$. 
Therefore, we can interpret the algebra of Poisson brackets for our 
model as corresponding to an infinite set of harmonic oscillators, 
described by the annihilation and creation like variables $a_k$ and 
$a_{k}^{\star}$, respectively. 

The quantization of this algebra can be performed by standard 
methods. For instance, one can simply introduce a Fock 
representation\cite{GJ}. In that case, our phase-space variables are
represented as annihilation and creation operators, $\hat{a}_k$ and 
$\hat{a}_k^{\star}$, and the Hilbert space of physical states can be
constructed by the repeated action of the creation operators on a 
vacuum, which is destroyed by all the operators $\hat{a}_k$.
The algebra of commutators is given by the quantum analog of 
Eq. (\ref{pbr}), 
\begin{equation} 
[\hat{a}_{\bar{k}},\hat{a}^{\star}_{k}]=\delta(\bar{k}-k),
\end{equation}
where we have set $\hbar=1$, and $\bar{k},k\in I\!\!\!\,R^+$. 
Assuming that the vacuum has unit norm, the inner product is totally 
determined by requiring that the relations under complex conjugation
between the variables $a_k$ and $a_k^{\star}$ are realized in the 
quantum theory as adjointness relations between operators. In 
particular, it is then possible to find normalized states which are 
formally similar to the states with $n$ particles or to the coherent
states of ordinary quantum field theory in flat spacetime.

In this way, one attains a consistent and well-defined mathematical
framework for analizying the quantum physics of spacetimes with two
commuting Killing vectors that describe pure gravitational plane 
waves with linear polarization.

\section{Conclusions and further comments}
\setcounter{equation}{0}

Pure gravitational plane waves have been considered as a special 
class of spacetimes with two commuting spacelike Killing vector 
fields. We have discussed the structure of the reduced phase space 
and the quantization of this family of solutions to vacuum general 
relativity.

We have first seen that, with a suitable choice of coordinates, the 
metric for these plane waves can be cast in a $3+1$ form in which the
values of the lapse and shift are fixed and all metric functions 
depend just on a coordinate of the sections of constant time, 
namely, the coordinate $u\in I\!\!\!\,R$. The other two coordinates 
of these sections (which run over the real axis) correspond 
to Killing vectors, with group orbits that are orthogonal to the 
surface described by $u$ and the time coordinate. We have shown
that the determinant of the metric for these group orbits can be 
chosen as a strictly increasing function of the coordinate $u$, with
range given by the interval $(0,1)$. In addition, we have solved 
explicitly Einstein equations in vacuum, and introduced metric 
variables whose reality ensures that the induced metric is positive 
definite. The family of metrics that we have obtained in this manner
represent the most general plane wave solution in source-free 
gravity, with the only caveat that they do not describe the whole 
spacetime which can in principle be covered with harmonic 
coordinates. These metrics are determined by two arbitrary functions
of the coordinate $u$, i.e., the functions $v$ and $y$.

With the aim at studying this family of metrics, we have considered
the Hamiltonian formalism for spacetimes that admit two commuting 
spacelike Killing vectors. Assuming the spatial topology to be that 
of $I\!\!\!\,R^3$, it has been proved that the condition that 
there exists a surface orthogonal to the group orbits removes
the gauge freedom related with diffeomorphims of the 
coordinates which correspond to the Killing vectors. We have 
then particularized our analysis to the case of pure gravitational 
plane waves. Using that the determinant of the metric for the group 
orbits is an increasing function of the coordinate $u$, we have 
been able to eliminate the degrees of freedom associated with 
diffeomorphisms of the $u$-line. In order to complete the gauge 
fixing and guarantee that the classical solutions of the system are 
plane waves, we have next introduced two types of requirements. On 
the one hand, we have imposed the symmetry conditions (\ref{X2}), 
which can be interpreted as relations between the metric
and the extrinsic curvature that are satisfied for plane waves. Once
these conditions are included, the densitized Hamiltonian constraint,
which is the only constraint that still remains on the system, is
easily solved. Moreover, all metric variables are then 
classically time independent, except possibly the variable $w$, 
defined by means of Eq. (\ref{dsgf}). On the other hand, we have 
demanded that $w$ adopts the same expression in terms of $v$ and $y$
as it does in pure gravitational plane waves. We have shown that, 
in this way, the gauge freedom associated with the densitized 
Hamiltonian constraint is totally removed and the time dependence 
of $w$ is ruled out on classical solutions. 

After this process of gauge fixing and symmetry reduction, we have 
arrived at a midisuperspace model which is free of constraints and 
whose line element coincides formally with that for plane waves in 
vacuum general relativity, although the two functions $v$ and $y$ 
that determine the metric may depend not just on the coordinate $u$,
but also on time if non-classical trajectories are allowed. The 
phase space of this reduced model has only two degrees of freedom, 
which are described by the variables $v$ and $y$. These variables 
remain constant in time on all classical solutions, so that the 
dynamical evolution on phase space is given by the identity map. 
This implies that the reduced Hamiltonian vanishes in the set 
of coordinates employed. The same conclusion has been reached by 
computing the reduced Hilbert-Einstein action. Up to surface 
contributions on sections of constant time, this action is linear 
in time derivatives, so that it indeed corresponds to a system with
vanishing Hamiltonian. Since the dynamics is trivial, all the 
information needed for quantizing the reduced model is provided by
its symplectic structure. Starting with the symplectic form 
for spacetimes with two spacelike Killing vectors, and 
calculating the pull-back to the section of the constraint surface 
determined by the gauge-fixing and symmetry conditions imposed on 
our model, we have then obtained the symplectic form on the reduced 
phase space. 

There exists at least a case in which the algebra of Poisson 
brackets that follows from this symplectic form is simple enough as 
to allow a straightforward quantization, namely, the case of 
linearly polarized plane waves. The restriction to this subfamily of 
plane waves in source-free gravity has been achieved by demanding 
that the function $v$ vanishes. This requirement can be viewed as 
an additional symmetry condition on the system. The phase space of 
the resulting reduced model is described by the metric 
variable $y$ or, equivalently, by the set of variables 
$\{a_k, a^{\star}_k;\;k\in I\!\!\!\,R^+\}$. These variables are 
classical constants of motion obtained from
the Fourier transform of the product of 
$y$ and a fixed function of the coordinate $u$. We have finally 
proved that the algebra of Poisson brackets for the reduced model 
of linearly polarized plane waves can actually be understood as
corresponding to an infinite set of harmonic oscillators whose 
annihilation and creation like variables are $a_k$ and $a^{\star}_k$,
respectively. The quantization of the model can then be readily 
performed by introducing a Fock representation for these variables. 

The mathematical framework constructed in this manner can be used to
study the quantum physics of pure gravitational plane waves with 
linear polarization. In particular, one can try to define operators 
for the metric components and analyze the quantum fluctuations of
the geometry. Another appealing possibility is to discuss the role 
played in the quantum theory by coherent states and see whether 
these are peaked around classical solutions, comparing the results 
with those obtained for linearly polarized waves with cylindrical 
symmetry\cite{AP,ASH97}. The consideration of these issues will be 
the subject of a future research.

Another direction for further investigation consists in introducing 
a scalar field in the model and studying its interaction with 
gravitational plane waves that are linearly polarized. This 
interaction has already been discussed in the general case of waves 
with two polarizations using a semiclassical approximation, i.e.,
considering only the quantization of the scalar field\cite{VER93}. 
It would be interesting to investigate 
whether the conclusions reached with this semiclassical analysis are
valid to some extent in a purely quantum theory.

Let us conclude with a couple of remarks about our reduced model 
for linearly polarized plane waves. The first comment refers to the 
algebra of Poisson brackets that we have found. It is not difficult 
to see that, at a given instant of time, the values of the metric 
variable $y$ at two different points do not generally commute under 
Poisson brackets. This implies that, in the quantum theory, there 
cannot exist states in which the variable $y$ takes a well-defined 
value at every point on a section of constant time, even though 
the points on this section are spacelike separated. In fact, this 
result is not so surprising. We have seen that the reduced 
Hamiltonian of the system vanishes in the set of coordinates 
adopted, so that the quantum evolution is dictated by the identity 
operator. Therefore, if it were possible to determine completely the
value of $y$ on a section of constant time, the same would happen 
for the entire spacetime, including points with timelike separation.
The second of our remarks concerns the vanishing of the reduced 
Hamiltonian. This conclusion is valid in the set of coordinates
that we have employed to describe the spacetime. In another 
coordinate system, however, the Hamiltonian of our reduced model can
generally differ from zero. Suppose, for instance, that we change 
coordinates from $u$ to $x=u-t$. Using the expression for the reduced
action given by Eq. (\ref{rlp}) in the linearly polarized case, one
can prove that the reduced Hamiltonian of the system becomes then
\begin{equation} 
\int dx\, (\partial_{x}Y)^2=\int_0^{\infty}dk\, ka_ka^{\star}_k,
\end{equation} 
where $Y$ is the variable defined in Eq. (\ref{Ylp}), and we have 
substituted relation (\ref{ft}) after replacing the coordinate $u$ 
with $x$. In the new coordinate system, the interpretation of our 
reduced model as a collection of harmonic oscillators applies hence 
not only to the Poisson algebra, but also to the Hamiltonian that 
generates the time evolution. 

\acknowledgments

The authors are grateful to L. J. Garay and C. Anastopoulos for 
valuable discussions. They wish to thank also E. Verdaguer, who
initially proposed the subject. G. A. M. M. acknowledges DGICYT 
for financial support under the Research Project No. PB94-0107. M. 
M. was supported by funds provided by a Basque Government FPI grant.

\end{document}